# Finding Significant Subregions in Large Image Databases

**Vishwakarma Singh**
`vsingh@cs.ucsb.edu`
Dept. of Computer Science,
University of California, Santa Barbara
Santa Barbara, CA 93106, USA.

**Arnab Bhattacharya**
`arnabb@iitk.ac.in`
Dept. of Computer Science and Engineering,
Indian Institute of Technology, Kanpur
Kanpur, UP 208016, India.

**Ambuj K. Singh**
`ambuj@cs.ucsb.edu`
Dept. of Computer Science,
University of California, Santa Barbara
Santa Barbara, CA 93106, USA.

## Abstract

Images have become an important data source in many scientific and commercial domains. Analysis and exploration of image collections often requires the retrieval of the best subregions matching a given query. The support of such content-based retrieval requires not only the formulation of an appropriate scoring function for defining relevant subregions but also the design of new access methods that can scale to large databases. In this paper, we propose a solution to this problem of querying significant image subregions. We design a scoring scheme to measure the similarity of subregions. Our similarity measure extends to any image descriptor. All the images are tiled and each alignment of the query and a database image produces a tile score matrix. We show that the problem of finding the best connected subregion from this matrix is NP-hard and develop a dynamic programming heuristic. With this heuristic, we develop two index based scalable search strategies, TARS and SPARS, to query patterns in a large image repository. These strategies are general enough to work with other scoring schemes and heuristics. Experimental results on real image datasets show that TARS saves more than 87% query time on small queries, and SPARS saves up to 52% query time on large queries as compared to linear search. Qualitative tests on synthetic and real datasets achieve precision of more than 80%.

**Category:** H.2.4[Information Storage and Retrieval]: Multimedia Search and Retrieval

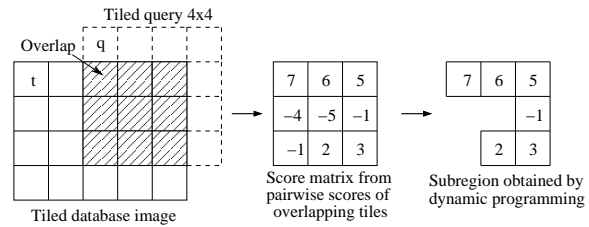

Figure 1: A $4 \times 4$ query is overlapped with a database image. For each tile in the $3 \times 3$ overlapped region, a score for the match is computed. Dynamic programming is run on the score matrix to obtain the maximal scoring connected subregion.

## 1 Motivation

Any image repository, be it a photo sharing site, a biomedical image database, an aerial photo archive or a surveillance system, must support retrieval, analysis, comparison and mining of images in order to be more useful than an online file cabinet. Over the years, image datasets have gained prominence in scientific explorations due to their ability to reveal spatial information and relationships not immediately available from other data sources. Decreasing cost of infrastructure and emergence of new tools have brought multimedia-based social networking sites and image-based online shopping sites into prominence. Existing query tools are being used in aerial photography or in face recognition for automatic image annotations [18, 25]; in astronomical satellite images to accurately detect point sources and their intensities [14]; and in biology for mining interesting patterns of cell and tissue behavior [4, 10]. The development of functionally enhanced, efficient and scalable techniques for image retrieval and analysis has the potential to accelerate research in these domains and open new frontiers for commercial and scientific endeavors.

Two problems have plagued the utility of content-based querying of multimedia databases: (i) the definition of similarity between images, and (ii) the restriction of distance measurement over semantically meaningful pairs of objects and sub-images. With regard to the former, a number of distance measures have been proposed such as Mahalanobis distance [19], earth mover's distance [23], and even learning the appropriate distance measure has been proposed [24]. The eventual goal is to identify a significant match for a pattern of interest in a query image. With regard to the latter problem, segmentation has been used to define the meaning and context of objects and segmentation-based image retrieval has been proposed. However, questions about the scalability of these techniques remain. We need to develop a technique that can find a meaningful subregion without much user intervention or extensive image analysis.

We propose a solution to pattern retrieval over images that extracts meaningful sub-images for a given query using the idea of a connected subregion. In order to define the best matching subregion, the idea of a score is adopted instead of measuring the pairwise distance. An appropriate scoring function is able to discriminate between foreground and background and is more amenable to the concept of subregions. Finally, images are decomposed into tiles with minimal amount of domain information to compute the score.

The idea of a match is illustrated in Figure 1. A query is aligned with each image in the database under all possible translations. Each alignment generates a matrix of scores, both positive and negative, between corresponding tiles. Positive scores denote foreground matches while a negative score means that a background tile of the query is matched to a database tile. A connected subregion over the matrix identifies a matching subregion. Scores over all possible connected subregions can be used to define answers to range and nearest-neighbor queries. The generality of the solution, the use of minimal domain knowledge, and the identification of best subregions are the unique aspects of our design.

Once we adopt the score and subregion based idea for retrieval of high-quality answers, the next challenge is one of scalability. How to identify the best subregions over millions of alignments? Clearly, a region-by-region search design is not going to work. How to develop access methods and index structures that can find the best subregions without examining all of them? Our solution to the scalability problem is two-fold: (i) development of an index structure that works with our definition of score, and (ii) design of two new algorithms that use the index structure to find the best subregions in an efficient manner.

The idea of finding the best connected subregion in a matrix that maximizes the sum of piecemeal scores is itself of theoretical and practical interest. We show that this problem is NP-hard. This necessitates appropriate heuristics that examine not all but a subset of connected subregions. We develop a dynamic programming based solution and characterize the class of subregions that is examined by this heuristic. Our access methods are unaffected by how the best connected subregions in an alignment are identified; they work correctly with any such heuristic.

In a nutshell, our contributions in this paper are as follows:

- We develop a score-based framework for identifying the best sub-images for a given query. Along with a tile-based decomposition of images and the identification of connected subregions, this technique is able to identify the best matching subregions for a given query (Section 2).

- We develop new access methods that use an index structure to identify the best matching subregions without exploring all images. The first method TARS is instance optimal but traverses the index multiple times and, therefore, performs better for small queries. The second method SPARS makes a single pass through the index and is suited for large queries (Section 3).

- We study the computational complexity of finding the highest scoring subregion. We show that this problem is NP-hard by reduction from the Thumbnail Rectilinear Steiner Tree problem [11]. We develop an efficient dynamic programming heuristic and characterize the class of subregions explored by this heuristic (Section 2).

- We save more than 87% of query time using TARS on small queries and more than 52% using SPARS on large queries as compared to linear search on two real large image datasets (Section 4). Extensive quality analysis on one synthetic dataset and three real datasets shows that our method returns significant results with precision of more than 80% (Section 4.4).

## 2 Computing Similarity using Scoring Function: Definition, Complexity, and Heuristics

In this section, we first discuss how the similarity between two image tiles is measured and how the idea of similarity is extended to regions. Then, we show that the computation of the optimal score between two images is NP-hard.

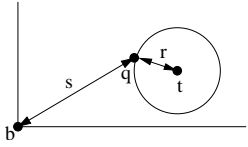

Figure 2: Scoring a query tile **q** against a database tile **t**. **b** denotes the perfect "background" tile. **score(q, t) = s − λr − c**.

Finally, we develop a dynamic programming heuristic to compute a good alignment. We search for similar regions for a given query image in a feature vector space. We split all the images in database into tiles of equal size. Our database $DB$ consists of the feature vectors of these tiles. We also tile the query image $Q$ and obtain feature vectors similar to a database image tile.

## 2.1 Scoring Function

We measure similarity between a pair of tiles using a scoring function. Our scoring function is a monotonically decreasing function of the distance between the feature vectors of a query tile $q$ and an image tile $t$. We can choose any image descriptor to transform an image tile into feature vector. The scoring function is defined as

$$score(q,t) = f(q) - g(d(q,t)) - c \quad (1)$$

where $f$ is a function based on domain knowledge, $g$ is a monotonically increasing function, $d$ is the distance and $c$ is a constant. The score can be positive or negative. We next explain the scoring function.

The intent of the scoring function is to discriminate between foreground (region of interest) and background. A query tile with little or no information forms the background and, therefore, should get a negative or low score no matter how good the match. A tile with more pattern information is a part of *region of interest* (ROI) and should get a high score when matched with a similar database image tile. The function $f(q)$ measures whether the tile $q$ is in a ROI.

The above function template is broadly applicable to a number of scoring functions. Next, we give a specific instance based on the idea of *log-odds* for the purpose of completion and experimental evaluation. In order to develop this model, we assume that the tile space consists of two distributions. The first distribution is that of foreground tiles from database which we call the *true distribution*. We model this as an exponential distribution[1]. For a database tile $t$ and a query tile $q$ (Figure 2), if $r = d(q,t) =$ the distance between the feature values of $q$ and $t$, then we can characterize this distribution as

$$P(q|true\ distribution) = \lambda_1 e^{-\lambda_1 r}. \quad (2)$$

The second distribution is that of background tiles in the database, which we call the *background distribution*. We postulate a perfect background tile $b$ (Figure 2) and an exponential background distribution centered at $b$. If $s = d(q,b)$, then

$$P(q|background\ distribution) = \lambda_2 e^{-\lambda_2 s}. \quad (3)$$

The score of a query tile $q$ matching a database tile $t$ is given by the *log-odds ratio*:

$$\begin{aligned} score(q,t) &= ln\frac{P(q|true\ distribution)}{P(q|background\ distribution)} \\ &= \lambda_2 s - \lambda_1 r + ln(\lambda_1/\lambda_2) \quad (4) \end{aligned}$$

Since scoring is only used to discriminate between foreground and background matches and the actual value is not important, the scores can be conveniently translated and scaled with constants. Denoting $\lambda_1/\lambda_2$ by a constant $\lambda$, then scaling by $\lambda_2$, and finally translating the score gives

$$\begin{aligned} score(q,t) &= s - \lambda.r - c \quad (5) \\ &= d(q,b) - \lambda.d(q,t) - c \quad (6) \end{aligned}$$

where $\lambda$ and $c$ are independent constants. Comparing Eq. (6) with Eq. (1), we see that $f(q) = d(q,b)$ and $g(x) = \lambda x$.

We name the above scoring function *discriminator function*. We can make the following observations from Eq. (6): (i) When distance to a database tile is kept invariant, a query tile with less background has a higher score, (ii) For a particular query tile, a more distant database tile has a lower score. We show the advantage of this scoring function over simple distance measures for similarity in Section 2.4.

The database querying algorithms (presented in Section 3) are independent of the scoring function as long as *the score is monotonically decreasing with the distance*.

## 2.2 Score of an Overlapping Region

Once we have a model to measure the similarity between a pair of tiles, we next consider how to measure similarity between two regions.

---

[1] The reasons for choosing this distribution are three-fold: first, observation from data, second, its simplicity, and third, its utility in capturing small variations over related images.

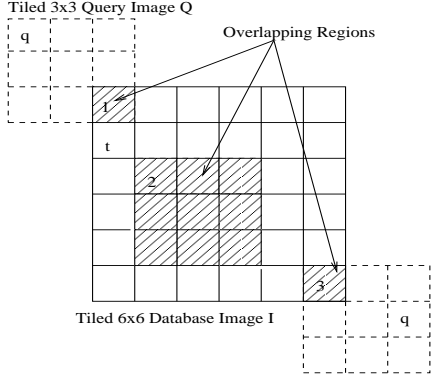

Figure 3: Overlapping regions found by translation of a query image $Q$ on a database image $I$ at 3 alignments.

The alignment or the overlap of a query image $Q$ with a database image produces a score matrix of pairwise aligned tiles, as depicted in Figure 1. The score of the alignment is defined as the score of a *connected subregion* that has the *maximal* possible *cumulative* score. We are interested in the alignment of a single pattern and, hence, the justification for finding a single connected region. The maximal scoring subregion may include negative scores and may not be rectangular in shape, as shown in Figure 1.

The best match in a database of images is found by considering all possible alignments, i.e., translations of the query image over each database image[2]. This is illustrated in Figure 3 where three alignments are shown.

## 2.3 NP-Completeness Proof

We next prove that the problem of finding the maximal scoring subregion in a score matrix is NP-hard. We prove this by showing that the corresponding decision problem is NP-complete. We first define the "graph" analog of the matrix problem as follows: *Given a graph representation $G = (V, E)$ of a matrix, with weight $w(v)$ on each vertex $v \in V$ corresponding to the entry in the matrix, is there a connected subgraph of weight $\geq W$?* We denote this problem by MAXIMAL WEIGHTED CONNECTED SUBGRAPH or MWCS.

**Theorem 1.** MAXIMAL WEIGHTED CONNECTED SUBGRAPH (MWCS) *is NP-complete, for a matrix graph of degree at most 4.*

*Proof.* MWCS is in NP since the weight of a connected subgraph can be computed in polynomial time.

---
[2]Rotations and reflections can be addressed by rotating and reflecting each alignment; however, they do not add to the technical development of the ideas and not considered further.

For our reduction, we use the RECTILINEAR STEINER TREE (RST) problem that is known to be NP-complete [12]. The RST problem asks: Given a set of $n$ terminal points that are embedded in an integer grid in a plane, is there a spanning tree of total length at most $l$ such that the vertices of the spanning tree are the input points of the set and the grid points, where the length of an edge is the $L_1$ distance between the corresponding vertices?

There is a special case of the RST problem known in the literature as the THUMBNAIL RECTILINEAR STEINER TREE (TRST) [11] problem. The TRST problem restricts the terminal points to an $m \times m$ grid. The TRST problem remains NP-complete even when $m$ is bounded by a polynomial of $n$ [12].

Given an instance of the TRST, we construct an instance of the MWCS as follows: We first find the bounding box of the points of the TRST, i.e., the $m \times m$ grid. Then, we replace each terminal point by a vertex of weight $w \gg l$. At each grid point that is not already occupied by the $n$ terminal points, we place a vertex with weight $0$. Between a pair of consecutive vertices on the same grid line (e.g., on the half-grid positions), we place a vertex with weight $-1$. Each vertex is connected to only to its horizontal and vertical neighbors, thus producing a matrix graph. Figure 4 shows an example of the construction. The original points are shown by double circles. The construction takes time polynomial in $m$, and hence polynomial in $n$, and the graph $G$ thus constructed is planar with degree at most 4.

We claim that the original TRST on $n$ points has a rectilinear Steiner tree of length $\leq l$ if and only if the MWCS graph has a connected subgraph of weight $\geq W = n.w - l$.

*Only if:* Assume that there is a Steiner tree of length at most $l$. By definition, it spans all the terminal points and is connected. Note that for a length $l$ path between two points, there are exactly $l$ vertices of weight $-1$. The vertices corresponding to the $n$ terminal points have a weight of $w$ each. Therefore, the weight of this tree is at least $n.w - l$. Figure 4 shows such a Steiner tree in solid lines.

*If:* Any connected subgraph of weight at least $n.w - l$ in $G$ must include all the $n$ vertices of weight $w$ and at most $l$ vertices of weight $-1$. There is no way to connect two vertices of weight $\geq 0$ without passing through a vertex of weight $-1$. Therefore, the length of this path is at most $l$, since otherwise, the connected subgraph would have included more than $l$ vertices of weight $-1$. Also, if the subgraph has the maximal weight, it is a tree, since, if it is not, at least one pair of vertices has more than one path between them. Removing that path increases the weight of the tree by the absolute weight of the negatively-weighted vertices in the path. Therefore, this subgraph defines a Steiner tree for the original $n$ points. An example of such a subgraph is shown in Figure 4 in solid lines. □

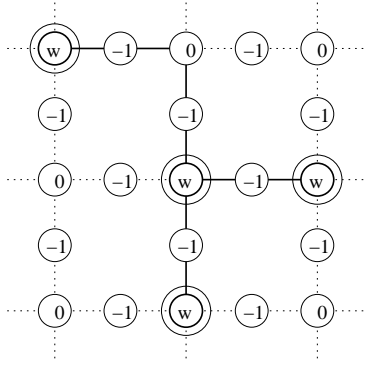

Figure 4: Construction from Thumbnail Rectilinear Steiner Tree instance to Maximal Weighted Connected Subgraph (MWCS) instance. The double lined vertices are the original terminal points. The bold lines represent the optimal solution of both the problems.

### 2.4 Dynamic Programming Heuristic

In this section, we design a simple dynamic programming (DP) heuristic as an alternative to examining all possible subregions for finding the maximal score. Assume that the score in cell $C(i, j)$ of the score matrix is denoted by $s(i, j)$. The DP starts from one of the corner cells of the score matrix. For discussion purposes, assume that it starts at cell $C(0, 0)$ in Figure 5. Next, it proceeds by first moving towards the right ($\rightarrow$) and calculates a subregion corresponding to each cell in $0^\text{th}$ row. Then it goes to $1^\text{st}$ row $0^\text{th}$ cell $C(1, 0)$ by moving in the top ($\uparrow$) direction in the score matrix (akin to a row-scan order). DP completes this iteration when it reaches the top-most and right-most cell in the matrix. In Figure 5, this cell is $C(2, 2)$.

$R(i, j)$ is a maximal scoring sub-region that has its top-right corner at the cell $C(i, j)$. Suppose $s(i, j)$ denotes the score of $C(i, j)$ and $S(i, j)$ denotes the maximal score for the subregion $R(i, j)$. DP examines 4 possibilities to find the maximal score of $R(i, j)$: (i) the score of the cell itself, (ii) the score of the cell plus the maximal score for the bottom subregion, (iii) the score of the cell plus the maximal score for the left subregion, and (iv) the score of the cell plus the maximal scores for the bottom and the left subregions. Since the bottom and the left subregions can intersect, the score of the intersecting region should be subtracted from the cumulative scores of the two subregions so that it is not counted twice.

The DP algorithm computes the following recurrence relation to find all the subregions in the score matrix and their scores:

$$S(i,j) = \max \begin{cases} s(i,j) \\ s(i,j) + S(i, j-1) \\ s(i,j) + S(i-1, j) \\ s(i,j) + S(i, j-1) + S(i-1, j) \\ \quad - S\big(R(i, j-1) \cap R(i-1, j)\big) \end{cases}$$
(7)

The corresponding subregions maintained for the 4 cases are, respectively:

$$R(i,j) = \begin{cases} C(i,j) \\ C(i,j) \cup R(i, j-1) \\ C(i,j) \cup R(i-1, j) \\ C(i,j) \cup R(i, j-1) \cup R(i-1, j) \end{cases}$$
(8)

To improve the overall score, DP executes the above logic starting from all the 4 corner cells with the following combinations of moves: (i) Starting at bottom-left cell and moving in $\uparrow$ and $\rightarrow$ direction, (ii) Starting at bottom-right cell and moving in $\uparrow$ and $\leftarrow$ direction, (iii) Starting at top-left cell and moving in $\downarrow$ and $\rightarrow$ direction, (iv) Starting at top-right cell and moving in $\downarrow$ and $\leftarrow$ direction. It returns the subregion having the maximum score of all these 4 possibilities. Such a subregion explored by DP on a score matrix is illustrated by Figure 1.

**Running Time:** For a score matrix of size $m \times n$, for each cell, the DP computes the maximal score for the subregion ending at that cell. Calculating the scores for each cell requires finding an intersection of the largest scoring subregions on its bottom and left. This requires a running time of $O(mn)$ in the worst case. Thus, the total running time of the DP algorithm is $O(m^2n^2)$. For a particular score matrix, the DP needs to be run from all the 4 corners, which is constant. Thus, the worst case running time for the DP is *quadratic* in the size of the score matrix.

**Class of Subregions Examined:** The DP algorithm is a heuristic, and therefore, does not investigate all the possible connected subregions; it chooses the maximal scoring connected subregion from only a certain class of shapes. Next, we analyze the class of such shapes. Consider only right ($\rightarrow$) and top ($\uparrow$) moves starting at left-bottom corner. The maximal scoring subregion $R(i, j)$ for cell $C(i, j)$ will include another cell $C(i', j')$ only if $C(i', j')$ is included in either $R(i, j-1)$ or $R(i-1, j)$. Similarly, the subregions $R(i, j-1)$ and $R(i-1, j)$ contain only those cells that are towards the left and bottom of them. Therefore, by induction, if cell $C(i', j')$ is included in $R(i, j)$,

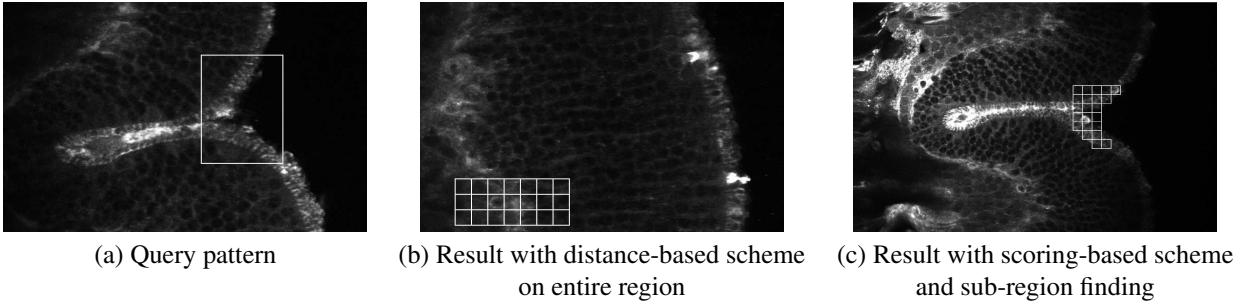

(a) Query pattern  (b) Result with distance-based scheme on entire region  (c) Result with scoring-based scheme and sub-region finding

Figure 7: (a) Example of a biologically interesting pattern. The marked pattern highlights a fold of the retinal tissue labeled with peanut-agglutinin conjugated to a fluorescent probe. (b) Retrieved result when distance-based matching on entire region is used. (c) Retrieved result when score-based matching on subregions is used.

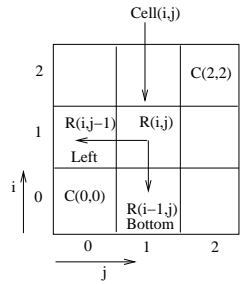

Figure 5: DP forms sub-region $R(i,j)$ by looking at scores of $C(i,j)$, $R(i-1,j)$ and $R(i,j-1)$.

| 3,1 (−1) | 3,2 (−1) | 3,3 (40) | 3,4 (−90) |
|---|---|---|---|
| 2,1 (−1) | 2,2 (10) | 2,3 (1) | 2,4 (35) |
| 1,1 (−1) | 1,2 (−1) | 1,3 (10) | 1,4 (−1) |

Figure 6: Example of shapes not captured by DP. The scores are shown in brackets. The optimal solution consists of the cells $(3,3)$, $(2,2)$, $(2,3)$, $(2,4)$ and $(1,3)$ having scores 40, 10, 1, 35 and 10 respectively.

then $C(i', j')$ must be at the left and bottom of $C(i, j)$. No cell which is towards the right or top of $C(i, j)$ will be included in $R(i, j)$. As an example, consider the cell $(2, 4)$ in Figure 6. It can consider only the cells $(i, j)$ where $i \leq 2$ and $j \leq 4$, i.e., all cells to its left and bottom. It cannot consider any other cell (e.g., cell $(3, 3)$ in the figure). The DP ends at the top-right corner. However, the maximal scoring subregion may end at any cell, and not necessarily at the top-right corner cell. Thus, for example, if the maximal scoring sub-region ends at $(3, 3)$, then the nature of DP forbids it to consider cells $(1, 4)$ and $(2, 4)$ (Figure 6). Hence, even though the optimal solution for the example in Figure 6 consists of all the five shaded cells, this DP will find only the region consisting of cells $(1, 3)$, $(2, 2)$, $(2, 3)$ and $(3, 3)$ as the answer. Shaded region given in Figure 6 can not be obtained by DP starting from any corner.

Since the DP is run from the four corners in four sets of moves, the subregions captured are of four types: (i) containing cells towards bottom and left, (ii) containing cells towards bottom and right, (iii) containing cells towards top and left, and (iv) containing cells towards top and right.

Formally, the shapes for the class of such subregions can be characterized in the following way. For a particular shape $P$, a cell $C(i, j)$ *sinks* another cell $C(i', j')$, denoted by $C(i, j) \triangleleft C(i', j')$, if $C(i, j)$ can be reached from $C(i', j')$ in $P$ by taking one of the four combinations of moves described earlier. For example, in Figure 6, cell $(3, 3)$ sinks cell $(2, 3)$ for right and top moves since it can be reached from $(2, 3)$ by this move combination. For the same move combination, it does not sink cell $(2, 4)$ as it cannot reached from $(2, 4)$ using only right and top moves. A cell $C(i, j)$ *sinks* a shape $P$, denoted by $C(i, j) \triangleleft P$, if and only if for all cells $C(i', j')$ belonging to $P$, $C(i, j)$ sinks $C(i', j')$, i.e.,

$$C(i,j) \triangleleft P \iff \forall C(i',j') \in P, \quad C(i,j) \triangleleft C(i',j') \tag{9}$$

A particular shape $P$ can be captured by DP if and only if there exists a cell $C(i, j) \in P$ that sinks $P$. Combining all the 4 sets of moves as mentioned earlier characterizes the entire set of shapes captured by DP. Examples of shapes captured are: $\sqcap, \square$, etc. Shapes that cannot be captured include $+, \times$.

**Advantages of Score Based Similarity:** We performed

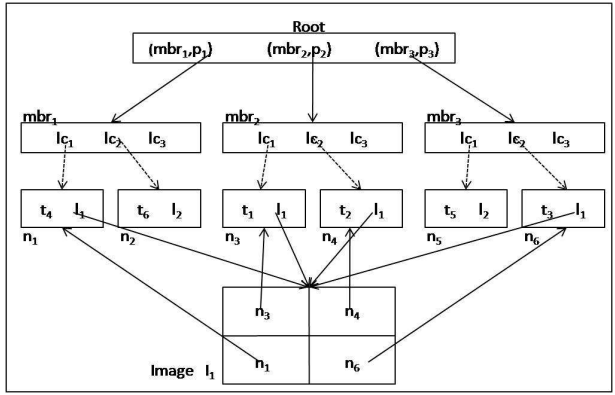

Figure 8: Index structure. Image $I_1$ maintains pointers to leaf nodes of its tiles. Leaf nodes maintain pointer to $I_1$.

quality experiments to compare our score based similarity measure with distance based measures. We used the *discriminator* scoring function (Section 2.1) to measure the similarity between a pair of tiles. We compared the first result retrieved by our similarity measure to a whole overlapping region $L_1$ similarity measure based on CSD like feature vector. One such result over biological images is shown in Figure 7. We can see that a simple distance measure fails to discriminate between foreground and background and hence generates more false matches. Our similarity measure maximizes the score of the best matching subregion and performs better.

## 3 Query Algorithms

In this section, we discuss linear search and two new query algorithms to find the top-$k$ similar regions from a database. The first algorithm is a naïve linear search through the database. The other two algorithms use a multi-dimensional index structure to prune the search space and achieve efficiency and scalability. In the ensuing discussion, the size of a query image $Q$ and a database image $I$ is defined in terms of the number of constituent tiles. We take the size of $Q$ to be $n$.

The *Linear Search* algorithm searches through all the possible overlaps to find the top-$k$ matching regions. It translates the query image over all the database images and computes a score for each of them. It maintains a priority queue of the results to find the $k$ highest scoring regions. Since the number of possible overlaps increases linearly with increase in image and database size, this method does not scale.

To make the search scalable and efficient, we next propose two algorithms TARS and SPARS. These algorithms use an index on the feature vectors of the image tiles to query nearest neighbors for a given tile. We can use any R-tree [13] (data-partitioning) like index structure for this. We choose bulk loadable STR-Tree [17] because of its simplicity and availability. Each leaf node is an entry of the form *lmbr(t,I)* where *t* is the feature vector of an image tile and *I* is a pointer to its parent image. Each non-leaf node is of the type *node(MBR,child-pointers)* where *MBR* is a minimum bounding box of the children and *child-pointers* are pointers to the child nodes. Each database image *I* is a two-dimensional array of pointers to the *lmbrs* containing its tiles as shown in Figure 8. This structure allows for full access from a tile to its parent image and vice versa. We can easily determine the row and column position of a tile from the image array to find an overlap.

**Algorithm 1** TARS
**Input:** $Q$: query image, $k$: number of matches
**Output:** $RQ$: priority queue of top-$k$ matches
1: $RQ \leftarrow [(\_, -\infty)]$
2: $T \leftarrow +\infty$
3: $BS$: bit-vector for explored overlap region
4: **while** $T \geq$ GetHead($RQ$).$s$ **do**
5:   **for all** $q_i \in Q$ **do**
6:     nnTile[i] $\leftarrow$ GetNextNN($q_i$)
7:   **end for**
8:   **for all** $q_i \in Q$ **do**
9:     $org \leftarrow$ FindOverlapRegion(nnTile[i],$q_i$)
10:     **if** $org$ not flagged in $BS$ **then**
11:       $sm \leftarrow$ GetScoreMatrix($Q$,$org$)
12:       $e(rg,s) \leftarrow$ DP($sm$)
13:       flag $org$ in $BS$
14:       Insert($RQ$, $e(rg,s)$)
15:     **end if**
16:     $sm \leftarrow$ score matrix of overlapping Q with nnTile
17:     $T \leftarrow$ score of DP($sm$)
18:   **end for**
19: **end while**

### 3.1 TARS (Threshold Algorithm for RegionBased Search)

The algorithm TARS formulates the region retrieval query as a top-$k$ aggregate query. It considers each tile $q_i \in Q, \forall i = 1, \ldots, n$ as an independent component of the query image. For each $q_i$, it takes the view that all database tiles are ranked in a decreasing order of their scores with $q_i$. Overlapping regions are determined by the tiles from the ranked lists. Table 1 shows a sorted view of the database for a query image consisting of 2 tiles.

| $q_1$ | $q_2$ |
|---|---|
| $(t_3,I_2,10)$ | $(t_1,I_1,9)$ |
| $(t_2,I_1,8)$ | $(t_2,I_1,7)$ |
| $(t_1,I_1,2)$ | $(t_3,I_2,6)$ |

Table 1: Sorted access of database tiles for a given query $(q_1, q_2)$ in TARS.

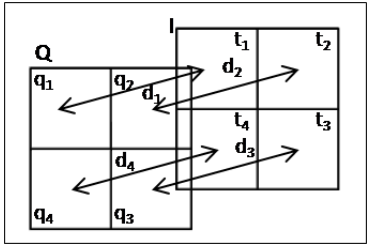

Figure 9: Overlap of query image $Q$ with database image $I$ such that $q_1$ aligns with $t_1$.

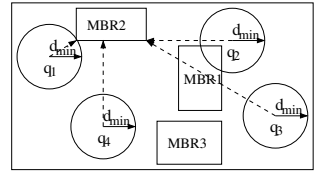

Figure 10: MBR and its nearest query tile. $q_1$ is nearest to $MBR_2$ with distance $d_{min}$.

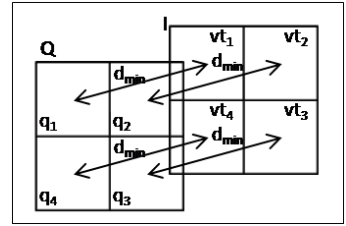

Figure 11: Overlap of query image $Q$ with virtual tiles $(vt_1, vt_2, \dots)$ at distance $d_{min}$.

Algorithm TARS adopts a strategy similar to the Threshold Algorithm (TA) [8] to solve the aggregate query. It performs incremental nearest-neighbor searches for each $q_i$ on the index structure to get sorted access to the database tiles. It starts by accessing the first nearest neighbor $t_{i_1}$ for each $q_i$ (steps 5-7 of Algorithm 1). Then, for each $q_i$, it finds the overlapping region of $Q$ with a database image $I$ ($t_{i_1} \in I$) such that $q_i$ aligns with $t_{i_1}$ in the overlap. Figure 9 shows how $Q$ having 4 tiles overlaps with an image $I$ also having 4 tiles such that $q_1$ aligns with $t_1$.

The algorithm then uses DP to find the maximum scoring subregion $rg$ and its score $s$ for each such overlap $org$ (steps 11-12). It inserts all the results $e(rg,s)$ in a priority queue $RQ$ of size at most $k$. The entries in $RQ$ are sorted based on their scores. Once $RQ$ has $k$ regions, a result is inserted only if its score is more than the current region with the least score. In order to prevent multiple processing of the same database region, TARS flags the explored overlapping regions in a bit-vector $BS$.

At the end of the first iteration, TARS builds a score matrix by aligning each $q_i$ with $t_{i_1}$. The DP score on this score matrix is the *threshold* score $T$. The threshold score is an upper bound on the scores of all the regions that have not been explored yet; this is because all the tiles to be accessed in the next iteration by each $q_i$ have scores lower or at best equal to the current $t_i$'s. This threshold score is updated after every iteration of the algorithm. The algorithm proceeds to the next iteration only if $T$ is greater than the least score in $RQ$. As TARS proceeds, $T$ decreases and the algorithm terminates eventually with optimal results.

The performance of algorithm TARS worsens with increase in query size. It is instance optimal but it traverses the index structure separately for each $q_i \in Q$ to access the database tiles in sorted order. The cost of this multiple nearest-neighbor traversal grows quickly with increasing query size. To avoid this scalability problem, we next propose a technique SPARS that finds the top-$k$ regions by performing a single traversal through the index structure and has better performance than TARS for large queries.

### 3.2 SPARS (Single Pass Region-Based Search)

Algorithm SPARS is a novel top-$k$ aggregate query algorithm which makes a single traversal through the index tree. It finds the top scoring regions by performing a *best-first search* [15]. It maintains a priority queue $BQ$ to find the next best node to process. When the algorithm encounters a leaf node, it explores an actual region in the database corresponding to its image tile. Similar to TARS, it maintains a priority queue $RQ$ of the top-$k$ regions.

The search for top-$k$ regions starts at the root node of the index, which is the first entry in $BQ$ with $+\infty$ score. The algorithm next processes each of its children. If the child is a non-leaf index node *mbr*, then it computes a score for it as follows (outlined in steps 8-12 of Algorithm 2). It determines the minimum distance between any query tile and the node $d_{min} = \min_i d(q_i, mbr)$, as shown in Figure 10. Then, it computes a score matrix by aligning each $q_i \in Q$ with *virtual* image tiles having a distance $d_{min}$ from $q_i$ as shown in Figure 11. The score

**Algorithm 2** SPARS

**Input:** $T$: tree root, $Q$: list of query tiles
**Input:** $k$: number of top results
**Output:** $RQ$: queue having top-$k$ results
1: $n \leftarrow \text{size}(Q)$
2: $RQ \leftarrow [(\_, -\infty)]$
3: $BQ$: queue of intermediate entities $\leftarrow [(T, +\infty)]$
4: $e(mbr,s) \leftarrow \text{GetHead}(BQ)$
5: **while** $e.s \geq \text{GetHead}(RQ).s$ **do**
6:   **if** $e$ is of type $(mbr,s)$ **then**
7:     **for all** child node $cn$ in $e.mbr$ **do**
8:       **if** $cn$ is $mbr$ **then**
9:         $d_{min} \leftarrow \text{GetMinDistance}(Q,mbr)$
10:         $sm \leftarrow \text{GetScoreMatrix}(Q,[d_{min}])$
11:         $e(rg,s) \leftarrow \text{DP}(sm)$
12:         $\text{Insert}(BQ,e(mbr,s))$
13:       **else**
14:         /*if $cn$ is a $lmbr$*/
15:         $q_j \leftarrow$ query tile nearest to $lmbr$
16:         $e(rg,s) \leftarrow \text{GetMaxSubRg}(lmbr,q_j,Q)$
17:         $RQ.\text{Insert}(e(rg,s))$
18:         **for all** $q_i$ in $Q$ and $i \neq j$ **do**
19:           $d_{min} \leftarrow \text{GetMinDistance}(q_i,lmbr)$
20:           $sm \leftarrow \text{GetScoreMatrix}(Q,[d_{min}])$
21:           $e(lmbr,q_i,s) \leftarrow \text{DP}(sm)$
22:           $\text{Insert}(BQ,e(lmbr,q_i,s))$
23:         **end for**
24:       **end if**
25:     **end for**
26:   **else**
27:     /*if $e$ is of type $(q_j,lmbr,s)$*/
28:     $e(rg,s) \leftarrow \text{GetMaxSubRg}(lmbr,q,Q)$
29:     $\text{Insert}(RQ,e(rg,s))$
30:   **end if**
31:   $e(mbr,s) \leftarrow \text{GetHead}(BQ)$
32: **end while**
33: **return** $RQ$

**Algorithm 3** Insert (SPARS)

**Input:** $PQ$: queue, $e$: entity
1: $e'(rg,s) \leftarrow \text{GetHead}(PQ)$
2: **if** $(\text{Size}(PQ) \geq k)$ **then**
3:   **if** $e'.s \leq e.s$ **then**
4:     **if** $PQ$ is $RQ$ **then**
5:       $PQ.\text{RemoveHead}()$
6:     **end if**
7:     $PQ.\text{insert}(e)$
8:   **end if**
9: **else**
10:   $PQ.\text{insert}(e)$
11: **end if**

**Algorithm 4** GetMaxSubRg (SPARS)

**Input:** $lmbr$: leaf node, $q_i$: query tile, $Q$: query tiles list
**Output:** $e$: entity of maxsubregion and score
1: $BS$: bit-vector for explored overlap region
2: $org \leftarrow \text{FindOverlapRegion}(lmbr,q_i)$
3: **if** $org$ not flagged in $BS$ **then**
4:   $dm \leftarrow \text{GetDistanceMatrix}(org)$
5:   $sm \leftarrow \text{GetScoreMatrix}(Q,dm)$
6:   $e(rg,s) \leftarrow \text{DP}(sm)$
7:   flag $org$ in $BS$
8: **end if**
9: **return** $e(rg,s)$

$s$ of the node is the score of the maximum scoring subregion found using DP on this score matrix. It inserts the *mbr* along with the score *s* as an element *e(mbr,s)* in *BQ*. It inserts $e$ into *RQ* only if *RQ* has less than $k$ elements or $s$ is greater than the minimum score in *RQ*.

If the child is a leaf node *lmbr*, then the algorithm finds the nearest query tile $q_i$ to tile $t$ of *lmbr* and computes the minimum distance $d_{min} = \min_i d(q_i, lmbr)$. It explores the actual image region for the $(q_i,t)$ alignment using Algorithm 4 as illustrated in steps 15-17 of Algorithm 2. Algorithm *GetMaxSubRg* finds the overlap of query $Q$ with image $I$ pointed by *lmbr* by aligning $q_i$ with $t$. Since the same overlapping region can be encountered later for a query and image tile pair, *GetMaxSubRg* maintains a bit-vector $BS$ to flag the explored regions; this prevents multiple processing of the same database region. Algorithm *GetMaxSubRg* returns the maximum scoring subregion *rg* of the overlap and the corresponding score *s* using DP. The result *e(rg,s)* is inserted in *RQ*.

After processing the alignment of $q_i$ with $t$, we still need to process the other alignments corresponding to other query tiles and $t$. The SPARS algorithm delays exploring these alignments in order to save computation cost. It calculates a score $s$ of the $lmbr$ for each $q_j, j \neq i$ using the same method discussed for a $mbr$ (outlined by the steps 18-25 of Algorithm 2). It finds the minimum distance $d_j$ between $q_j$ and tile $t$ in $lmbr$. It overlaps the query image with a virtual image such that the distance between each aligned pair of tiles is $d_j$. DP is run on this score matrix to compute score $s$ of the maximum scoring subregion for the overlap. SPARS inserts elements $e(lmbr,q_j,s)$ in *BQ* for each $q_j$ using. The algorithm explores these regions during the access of the elements from $BQ$ as outlined by steps 28-29 of Algorithm 2.

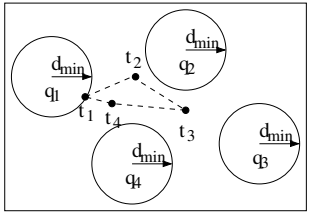

Figure 12: Tiles of the overlapping region for $q_1$ aligning with $t_1$ lie at distances greater than $d_{min}$.

The best-first search proceeds by accessing the current highest scoring element from *BQ*. SPARS terminates when the lowest score in *RQ* is greater than the highest score in *BQ*.

**Pruning Strategy:** The scoring function $s(q,t)$ is a monotonically decreasing function of $d(q,t)$, as discussed in Section 2.1. The aggregate score of an overlap is also monotonic with respect to individual scores of the score matrix of an overlap. With this monotonicity property, the following lemma holds. SPARS uses this lemma to prune the search space.

**Lemma 1.** *The score $S(Q,I)$ of the overlap of a query image $Q$ with an image $I$ with $d(q_i, t_i) = r$, $\forall i$, is an upper bound on the score $S(Q, I')$ of all possible overlaps of $Q$ with image $I'$ provided $d(q_i, t'_i) \geq r$, $\forall i$.*

From this lemma, we see that the score $s$ of a node (*mbr* or *lmbr*) at a minimum distance $d_{min}$ from the query tiles is an upper bound on the score of all nodes whose minimum distance is greater than $d_{min}$. We visualize such an example in Figure 10 where $MBR_2$ is at a distance of $d_{min}$ but $MBR_3$ is at a greater minimum distance from all the query tiles. Therefore, score of $MBR_3$ will be less than $MBR_2$. The score $s$ is also an upper bound on the score of an actual overlapping region if the distance between the corresponding tiles of $Q$ and $I$ have distance greater than $d_{min}$. We have such a scenario in Figure 12 in which tile $t_1$ finds $q_1$ as its nearest neighbor. All the tiles of the corresponding overlap as shown in Figure 9 lie at a distance greater than $d_{min}$. Therefore, the score of this overlap is less than the score $s$ of a node. These facts justify that the score of an element in *BQ* is an upper bound on all the nodes and regions that have not been explored and are ranked lower in *BQ*. At any point during the search, SPARS has already explored a hypersphere of radius $d_{min}$ centered at each query tile if the next candidate from *BQ* has a minimum distance $d_{min}$ from all the query tiles.

SPARS processes the nodes in decreasing order of their scores. It explores all nodes having score greater than the least score in *RQ* as they are potential candidates to yield regions with higher score. It terminates the search once the minimum score in *RQ* becomes more than the highest score in *BQ*. Thus, this pruning strategy ensures an optimal result for SPARS.

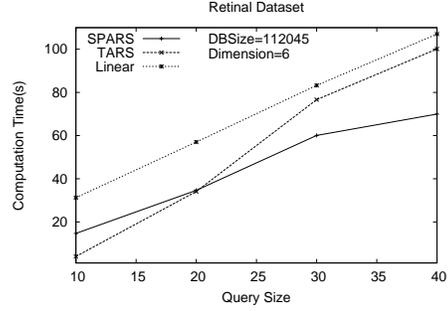

Figure 13: Effect of query size on the performance of the algorithms for retinal images.

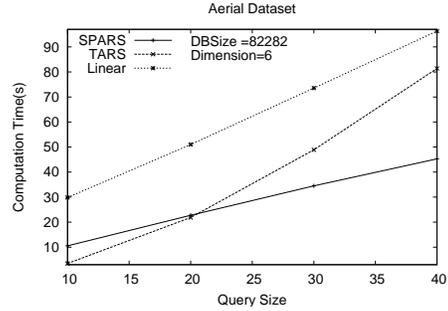

Figure 14: Effect of query size on the performance of the algorithms for aerial images.

## 4 Experimental Studies

In this section, we first empirically analyze the performance and efficiency of our access methods. Then, we present detailed quality analysis of our new similarity measure with visual results. We used Java 5.2 as our implementation language. We performed experiments on a 3.2GHz, 4GB memory PC running Debian Linux 4.0.

### 4.1 Dataset Preparation

We used two large real image datasets to empirically analyze the efficiency and scalability of our algorithms. The first dataset was 112,045 gray-scale images of various tissues and layers of retina [10] for different experimental conditions. Multiple molecular probes such as lectins and antibodies are used to examine the localizations of specific protein expression in retinal cells and the expression patterns of these proteins in different layers of retina. The fluorescence tagged probes are imaged by immunohistochemistry using confocal microscopes. We used the magnification of these images to scale all the images to a standard magnification using the CubicFilter from GraphicsMagick[3].

| Reduced dimension | Energy retained ||
|---|---|---|
| | Retinal dataset | Aerial dataset |
| 3 | 85.73% | 81.21% |
| 6 | 96.14% | 93.38% |
| 13 | 98.94% | 97.55% |

Table 2: Percentage energy remaining after dimensionality reduction (PCA).

Our second dataset was 82,282 gray-scale aerial images from the Alexandria Digital Library[4]. These are satellite images and air photos of different regions of California. The size of the images in both data sets varied from $320 \times 160$ pixels to $640 \times 480$ pixels.

We computed the feature vectors from the images in the following way. We split the images into non-overlapping tiles of size $32 \times 32$ pixels. We can use any image descriptor to transform image tiles into feature vectors. We choose to extract 256-dimensional feature vectors from each tile using a method similar to the Color Structure Descriptor (CSD) of MPEG-7 [21]. CSD is simple to compute and provides rotation invariant features that capture local structure in the image. To enhance efficiency, we performed PCA on these feature vectors to reduce the dimensionality. The number of principal components retained and the corresponding energy preserved is shown in Table 2. The index structure was built on this transformed data. We used the *discriminator* scoring function discussed in Section 2.1 to measure the similarity between a pair of tiles. We discuss the choice of parameters for the scoring function later in Section 4.4.

The parameters that are crucial to the performance of the access methods are: (i) Query size, $n$ (ii) Database size, $N$, and (iii) Dimensionality of the feature vector, $dim$. Query size $n$ is defined as the number of constituent tiles in the query image. Database size $N$ is defined as the number of images. The number of images and possible overlapping regions obtained by translation for varying $N$ is described in Table 3 for both retinal and aerial datasets. For each experiment, we used 100 randomly picked queries from the dataset. All the reported

---
[3] http://www.graphicsmagick.org/
[4] http://www.alexandria.ucsb.edu/

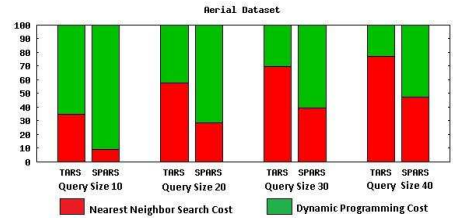

Figure 15: Percentage split of NN and DP time for varying query sizes for TARS and SPARS for aerial images.

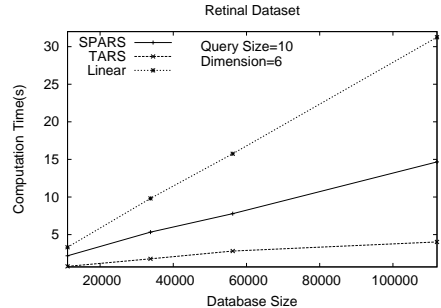

Figure 16: Performance of algorithms for varying database sizes of retinal images for query size 10.

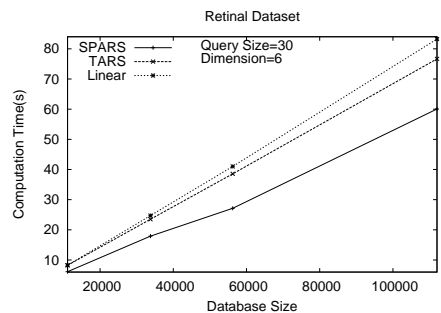

Figure 17: Performance of algorithms for varying database sizes of retinal images for query size 30.

time measurements are averaged over these 100 queries for top-10 results.

## 4.2 Performance Comparison of the Algorithms

We experimented with varying query sizes to compare the performance of the algorithms. We use the largest datasets of size $N = 112,045$ for retinal images and $N = 82,282$ for aerial images with $dim = 6$ for this experiment. Our results show that both TARS and SPARS outperformed *Linear Search* on both the datasets, as shown by Fig-ures 13 and 14. When we compared TARS with SPARS on the aerial dataset, we found TARS to be 3 times faster for query size 10 but slower by 2 times for query size of 40 than SPARS (Figure 14). The same behavior was noticed for the retinal dataset where TARS was faster by 3.6 times for $n = 10$ but slower by 1.4 times for $n = 40$ than SPARS (Figure 13).

| Retinal dataset | | Aerial dataset | |
|---|---|---|---|
| Number of images | Number of regions | Number of images | Number of regions |
| 112,045 | 10,004,850 | 82,282 | 10,625,200 |
| 56,241 | 5,000,050 | 37,037 | 5,000,000 |
| 33,762 | 3,000,000 | 21,744 | 3,000,000 |
| 11,112 | 1,000,100 | 5,560 | 1,000,000 |

Table 3: Database sizes of retinal and aerial images.

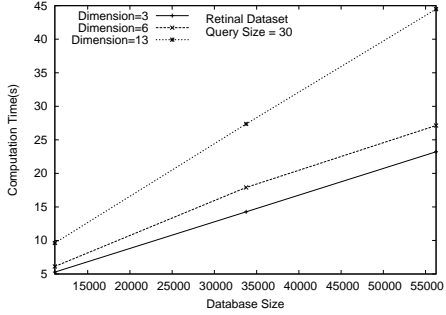

Figure 18: Effect of database size and dimension on the performance of SPARS on retinal images.

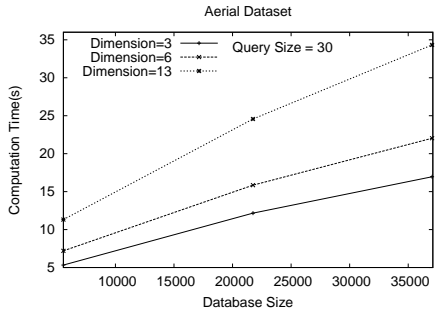

Figure 19: Effect of database size and dimension on the performance of SPARS on aerial images.

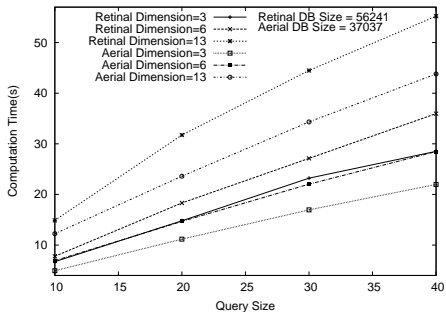

Figure 20: Effect of query size and dimension on the performance of SPARS on retinal and aerial images.

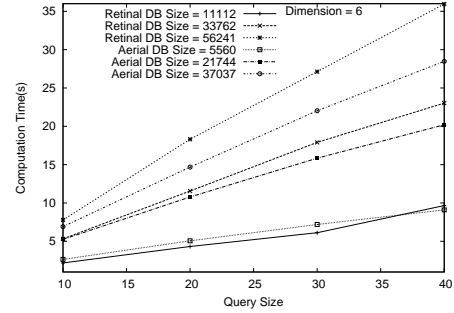

Figure 21: Performance of SPARS for varying query sizes and database sizes of retinal and aerial images.

SPARS performs better than TARS for query sizes of more than 20. We attribute this change in performance of TARS to its multiple traversal through the index structure, as discussed in Section 3.1. We measured the average of total nearest-neighbor search cost *NN* and dynamic programming *DP* cost for varying query sizes for both the algorithms TARS and SPARS. We present the percentage of time spent by each algorithm on NN and DP in Figure 15. We observe that the *NN* cost increases faster in TARS compared to SPARS as query size increases. TARS is instance optimal [8] and, therefore, it performs better than Linear and SPARS for smaller query sizes when the cost of this multiple traversal is not high. As this cost increases with increase in query size, its performs poorer than SPARS.

We next experimented with varying database sizes for the retinal dataset to confirm the above behavior of the algorithms. For a query size of $n = 10$ and $dim = 6$, we found TARS to be more than 2.7 times faster than SPARS across the database sizes as shown in Figure 16. SPARS is more than 1.5 times faster than the Linear

Search. The performance difference increases with increase in database size. Our other experiment with $n = 30$ and $dim = 6$ found SPARS to be 1.3 time better than TARS and more than 1.3 times better than Linear (Figure 17).

It is evident from the experimental results discussed above that TARS is a better algorithm than the other two for $n \leq 20$ whereas SPARS is better for $n > 20$. TARS saves more than 87% of the query time for $n = 10$ on both the datasets for the largest size. SPARS has a saving of 34% on retinal and 52% on aerial for $n = 40$ on the largest datasets. The average query time of TARS is approximately $4\,s$ on a database of size $N$ =112,045 and a query size of $n = 10$. The average query time for SPARS on the same database is $70\,s$ for query size of 40.

## 4.3 Performance Analysis of SPARS

We next performed detailed analysis of the behavior of the algorithm SPARS for varying $n$ (query size), $N$ (database size) and $dim$ (dimension) on both the datasets. The performance results of SPARS on both datasets for varying $N$ and $dim$ are shown in Figures 18 and 19. The dataset size is 56,241 for retinal images and 37,037 for aerial images. SPARS scales linearly for a given query size across varying database sizes and dimension. The performance of SPARS on both datasets for varying $n$ and $dim$ is shown in Figure 20. We found a linear behavior for a given database size across varying query sizes and dimensions. Experiments with varying $N$ and $n$ for $dim = 6$ on both datasets also revealed a linear performance as shown in Figure 21.

The exhaustive set of empirical results discussed above confirms a linear performance for SPARS across varying query size, database size and dimension. This establishes the scalability and efficiency of the algorithm.

## 4.4 Quality Analysis

In this section we analyze quality of our similarity measure and describe the datasets used for experiments.

**Dataset Preparation:** We used 4 different datasets to verify the quality of our new similarity semantic. From each dataset, we chose interesting regions for querying. For each query, regions in images were manually tagged as a true or a false match. Since the process is manually intensive, we used small datasets as shown in Table 4. PA and NF datasets are confocal microscopic images of cross-sections of feline retina labeled with the lectin peanut-agglutinin and anti-neurofilament antibody respectively. Aerial dataset consists of satellite images of Beverly Hills in California. We also created a synthetic dataset of 290 random images taken from the web. We segmented regions from images. Then, we created multiple images by placing a selected set of regions of interest in various environments. We used either a black or a gray background. A particular region was occluded to various degrees by another set of regions in different orientations. We use it to exhibit different characteristic and quality of our similarity measure.

**Parameter Learning and Precision:** We use pure black as background for retinal images which is true for the most of the real microscopic images. We use pure black for aerial images also for the purpose of simplicity, though, it can have other backgrounds. Background for other domains need to be determined from knowledge and training. We take the sum of all the pixels in a tile as its distance from background.

We learn the parameter $\lambda$ and $c$ using manual training with an approach similar to Walrus [22]. For each query, we measured $top$-$k$ precision where $k = 5$ and precision is the ratio of true matches to total matches. We trained the PA dataset on 10 queries. Highest precision (82.0%) was achieved for $\lambda = 1$ and $c = 23000$. These parameters gave an accuracy of 78.6% for 8 other queries over PA dataset. For the same parameter values, 8 queries on NF dataset and 10 queries on synthetic dataset gave precision of 82.5% and 100% respectively. Training and testing on 7 aerial image queries had a precision of 88% for $\lambda = 1$ and $c = 115000$. We summarize the results in Table 4. The same parameter values were used for scalability and efficiency measurements in Section 4.2. Our size of the training set was limited by the manually intensive nature of the task.

**Similarity Measure Characteristics:** We used the synthetic dataset to validate 3 novel characteristics of our similarity measure. First, it supports multi-region queries. As shown in Figure 22(a), the query extends over different regions of the input image. Second, the scoring function discriminates foreground from background and, therefore, discards background from the matched region. As shown in Figure 22(b), the matched tiles mostly lie in the region of interest (ROI). Third, it computes the best matching subregion of a given query with a region of an image rather than computing the overall similarity of the overlap. This property enables it to match partial and occluded regions. Our similarity measure retrieves Figure 22(b) as the first match which contains ROI partially and Figure 22(c,d) as the second and third match in which ROI is occluded by another region.

| Dataset | Images | Queries | $\lambda$ | $c$ | Precision |
|---|---|---|---|---|---|
| PA Retinal | 80 | 18 | 1 | 23000 | 80.3% |
| NF Retinal | 37 | 8 | 1 | 23000 | 82.5% |
| Aerial | 550 | 7 | 1 | 115000 | 88% |
| Synthetic | 290 | 10 | 1 | 23000 | 100% |

Table 4: Datasets used for quality analysis, corresponding parameter values for scoring function, and precision measures.

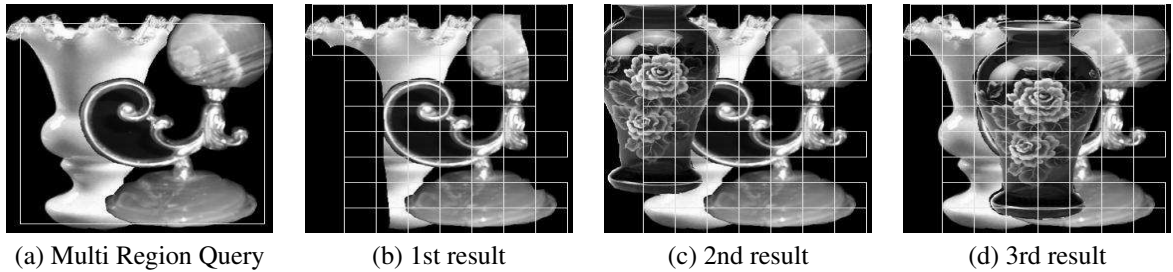

(a) Multi Region Query  (b) 1st result  (c) 2nd result  (d) 3rd result

Figure 22: (a) A multi-region query from synthetic dataset. (b) Algorithm retrieves image where objects are partially present. (c, d) Algorithm retrieves images where objects are occluded.

Finally, we present results for a set of queries from all the 4 datasets in Figure 23. All the match for retinal images have been validated by domain scientists and found significantly interesting. As shown in Figure 23(2), query and the results are examples of biologically interesting fold of retinal tissues labeled with peanut-agglutinin.

## 5 Related Work

Content-Based Image Retrieval (CBIR) systems either use a single global feature vector or some aggregation of various global features extracted per image to answer similarity queries. An excellent survey on the recent methods of CBIR can be found in [7]. Region-Based Image Retrieval (RBIR) systems extend CBIR by making the search sensitive to different regions of an image. It segments the image into a number of homogeneous regions and extract local features for each region. Similarity between a query and a database image is an aggregate function of the similarity between the associated query and database image regions. RBIR not only improves search quality but supports new query types based on the defined regions or objects in a query image.

Most of the RBIR systems use automatic or manual region segmentation in order to characterize regions and then compute a one-to-one or many-to-one mapping to match query regions to those in the database [1, 2, 5, 22, 26, 27]. Grids were used by [3] to partition images into clusters in the color space. Each image was modeled as a graph. The database candidates were filtered using a progressive distance measure with the query image graph. These segmentation-based RBIR systems are incapable of handling region queries that partially extend across various segments or regions of an image. Malki et al. [20] avoided segmentation by using a multi-resolution quadtree [9] to organize images. They used a normalized distance in each level of the quadtree and combined them for region queries. None of the above research discerns between foreground and background except [6], where local low-level features were used to remove irrelevant background. Further, most of the methods scan the database sequentially [1, 16]. As a result, they do not scale with database size. Weber et al. [27] used lower and upper bounds on distances between segmented regions as a filtering step to obtain candidate database regions that were then matched fully. Bartolini et al. [2] used an incremental nearest-neighbor algorithm to retrieve database matches for each query component. The matches were then combined using an aggregate function. However, they used segmentation to obtain the query results. Subregions containing patterns of interest were not identified.

## 6 Conclusions

In this paper, we addressed the problem of querying significant image subregions. We designed a generic scoring scheme to measure similarity between a query image and an image region which can use any kind of image descriptor. We also present an specific instance of the scoring function. We tiled the images to represent a re-gion as a collection of tiles, and each overlap between a query and a database image as a matrix of scores. We proved that the problem of finding a connected subregion of maximal score in a score matrix is NP-hard and then developed a dynamic programming heuristic to score an overlapping region. With this similarity measure, we proposed two index based scalable search strategies TARS and SPARS for querying in a large repository. We empirically analyzed the performance of these algorithms on datasets of 112,045 retinal images and 82,282 aerial images. We save more than 87% search time on small queries using TARS and up to 52% search time on large queries with SPARS on these datasets as compared to linear search. It should be noted that our heuristic for finding the best connected subregions and our access methods for top-$k$ queries (TARS and SPARS) are independent of each other. We demonstrate the quality of our similarity measure (more than 80% precision) with analysis over 3 real and 1 synthetic datasets. The ability to extract significant subregions (connected regions with highest score) can have a significant impact on analyzing growing collections of multimedia objects. Future work includes the formulation of other heuristics for finding similar subregions that have bounded approximation errors on quality and the formulation of other domain-specific scoring functions.

| Type | No. | Query Image | 1st Result | 2nd Result |
|---|---|---|---|---|
| Synthetic | 1 | 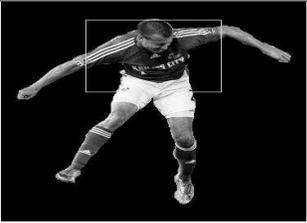 | 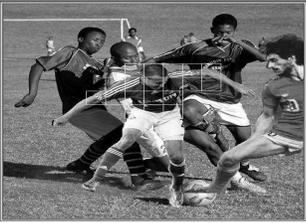 | 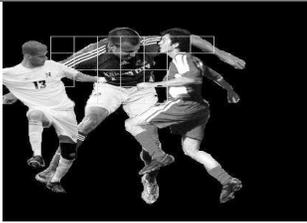 |
| PA Retinal | 2 | 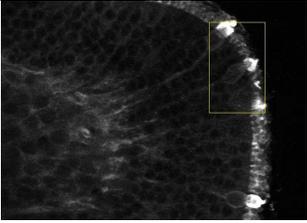 | 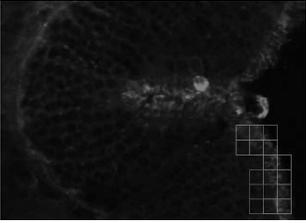 | 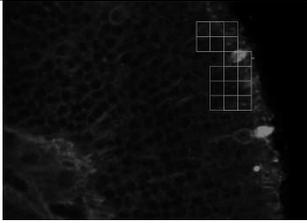 |
| | 3 | 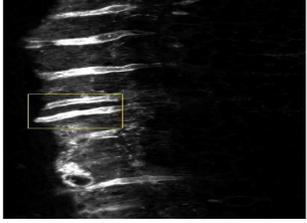 | 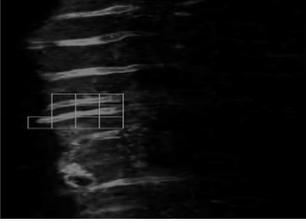 | 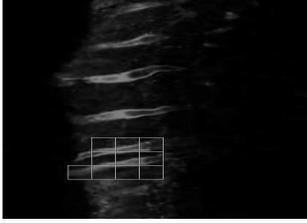 |
| NF Retinal | 4 | 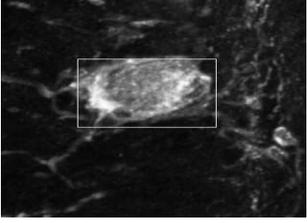 | 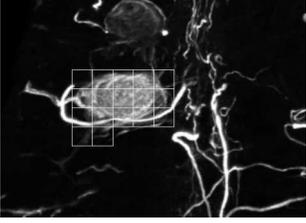 | 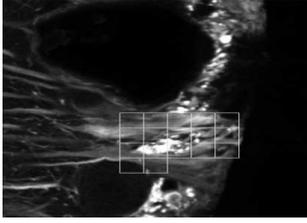 |
| | 5 | 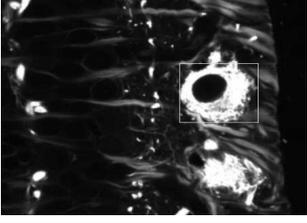 | 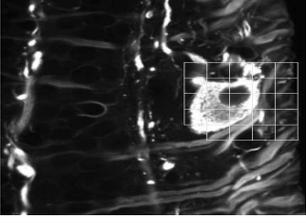 | 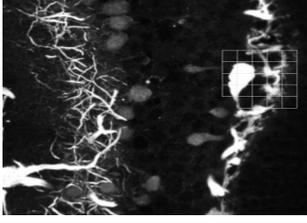 |
| Aerial | 6 | 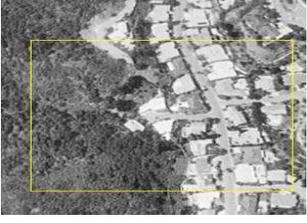 | 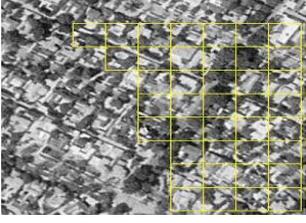 | 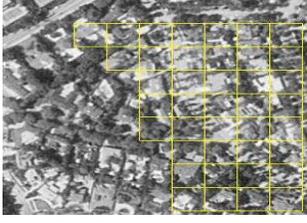 |
| | 7 | 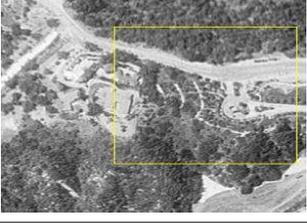 | 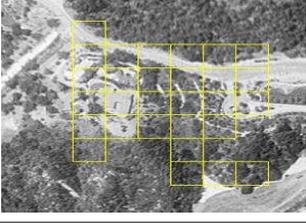 | 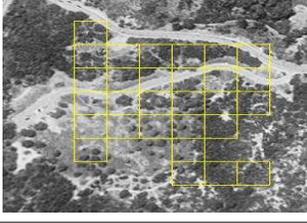 |

Figure 23: Top-2 results for various queries from different datasets.